\documentclass[11pt]{article}
\usepackage[a4paper,hmarginratio=1:1,vmarginratio=2:3,totalwidth=15.1cm,
totalheight=21.5cm]{geometry}
\usepackage{bm,epstopdf,epsfig,amsmath,amssymb,amsfonts,colordvi,
latexsym,comment,cancel,verbatim}
\usepackage[font=md,captionskip=8pt]{subfig}
\usepackage[usenames,dvipsnames]{color}

\renewcommand{\baselinestretch}{1.12}

\newcommand{\cD}{{\cal D}}

\newcommand{\cJ}{{\cal J}}

\newcommand{\cL}{{\cal L}}

\newcommand{\cO}{{\cal O}}

\newcommand{\ra}{\rightarrow}
\newcommand{\be}{\begin{equation}}
\newcommand{\ee}{\end{equation}}
\newcommand{\bea}{\begin{eqnarray}}
\newcommand{\eea}{\end{eqnarray}}

\newcommand{\otheta}{\overline\theta}
\newcommand{\opsi}{\overline\psi}

\long\def\symbolfootnote[#1]#2{\begingroup
\def\thefootnote{\fnsymbol{footnote}}\footnote[#1]{#2}\endgroup} 
\begin{document}

\begin{flushright}
CERN-PH-TH/2012-231\\
\today
\end{flushright}

\thispagestyle{empty}

\vspace{3cm}

\begin{center}

{\Large {\bf Low-scale SUSY breaking 
and the (s)goldstino physics.}}\\
\medskip
\vspace{1.cm}
\textbf{
 I. Antoniadis$^{\,a}$\,\symbolfootnote[2]{on leave from CPHT (UMR CNRS 7644) 
 Ecole Polytechnique, F-91128 Palaiseau, France.}, 
D.~M. Ghilencea$^{\,a,\,b}$
\symbolfootnote[3]{e-mail addresses: 
dumitru.ghilencea@cern.ch; ignatios.antoniadis@cern.ch.}} \\

\bigskip

$^a\,${\small CERN - Theory Division, CH-1211 Geneva 23, Switzerland.}\\

$^b\,${\small Theoretical Physics Department, 
National Institute of Physics }\\

{\small and Nuclear Engineering (IFIN-HH) Bucharest, MG-6 077125, 
Romania.}

\end{center}

\def\baselinestretch{1.14}
\begin{abstract}
\noindent
For a 4D N=1 supersymmetric model with a low SUSY breaking scale ($f$) and
general Kahler potential  $K(\Phi^i,\Phi_j^\dagger)$ and superpotential $W(\Phi^i)$  
we study, in an effective theory approach, the relation of the goldstino superfield 
to the (Ferrara-Zumino) superconformal symmetry breaking  chiral superfield $X$.   
In the presence of more sources of supersymmetry breaking, we verify the conjecture   
that the  goldstino superfield  is the (infrared) limit  of $X$ for zero-momentum
and  $\Lambda\ra \infty$ ($\Lambda$ is the  effective cut-off scale).
We then study the constraint $X^2=0$, which in the one-field case is known to decouple a 
massive sgoldstino and thus provide an effective  superfield description 
of the  Akulov-Volkov  action for the goldstino. In the presence of additional 
fields that  contribute to SUSY breaking we  identify conditions for which 
$X^2=0$ remains valid, in the effective theory below a large but finite 
sgoldstino mass. The conditions ensure that the effective expansion 
(in $1/\Lambda$)  of the initial Lagrangian is not in conflict with 
the decoupling limit of  the sgoldstino  ($1/m_{\rm sgoldstino}\sim \Lambda/f$, 
$f<\Lambda^2$).
\end{abstract}

\newpage

\section{Introduction}

Supersymmetry, if realised in Nature, must be broken at some high scale.
In this work we consider the case of SUSY breaking at a (low) scale $\sqrt f\!\ll \!M_{Planck}$
in the hidden  sector, an example of which is gauge mediation. In this case 
the transverse gravitino  couplings ($\sim 1/M_{Planck}$) can be neglected 
relative to their longitudinal counterparts ($\sim1/\sqrt f$) that  are due to its 
goldstino component. If so, one can then work in the gravity 
decoupled limit, with a massless  goldstino. The auxiliary field 
of the goldstino superfield breaks SUSY spontaneously,
while the goldstino scalar superpartner (sgoldstino) can acquire mass and decouple
 at low energy, similar to SM  superpartners, to leave a non-linear SUSY realisation.
To describe this regime one can work with component fields and
 integrate out explicitly the sgoldstino and other superpartners too (if massive),
 to obtain the effective Lagrangian. Alternatively one can use 
a less known but elegant superfield formalism  endowed with constraints,  
see \cite{SK} for a review. If applied to matter, gauge and goldstino
superfields, these constraints  project  out  the massive superpartners,
giving a superfield action for the  {\it light} states. Such constraint may be applied 
only to the goldstino superfield which can be coupled to the linear  multiplets of
the model (e.g. MSSM), to parametrize SUSY breaking.

In this work we  study SUSY breaking in the hidden sector and 
its relation to the goldstino  superfield in the presence of more sources of SUSY breaking
and  the connection of the goldstino superfield to the superconformal symmetry 
breaking chiral superfield  $X$~\cite{Ferrara}. 

It was noticed long ago  \cite{Casalbuoni:1988xh} (see also \cite{SK}) that a Lagrangian,
function of $\Phi^1=(\phi^1, \psi^1,F^1)$
\bea
\cL=\int d^4\theta \,\,\Phi_1^\dagger\,\Phi^1+\Big\{\int d^2\theta\, f\,\Phi^1 +h.c.\Big\},
\quad\mbox{with a constraint:} \quad (\Phi^1)^2=0,
\label{g}
\eea
provides an onshell superfield description of the Akulov-Volkov action for the 
goldstino field \cite{Volkov:1973ix}.
 Indeed, the constraint (which generates interactions)
has a solution $\Phi^1=\psi^1\psi^1/(2F^1)+\sqrt 2\theta\psi^1+\theta\theta F^1$, 
which ``projects out'' the sgoldstino $\phi^1$.  When this $\Phi^1$ is used back in 
(\ref{g}), one obtains the onshell-SUSY Lagrangian for a massless  goldstino $\psi^1$.
If more fields are present and for  general $K$, $W$,  the situation is more  complicated
and was little studied. 

Further, it was only recently conjectured \cite{SK} that the goldstino superfield is 
the infrared (i.e. zero-momentum)  limit of the superconformal  symmetry breaking  
chiral superfield $X$, that breaks the conservation of the Ferrara-Zumino current 
\cite{Ferrara}. In the light of the above discussion, 
one then also expects that in such limit $X^2\sim (\Phi^1)^2=0$, and 
this conjecture was verified in  very simple examples.  We address  
these two problems, more exactly:

\noindent
{\bf a)} the convergence of the field $X$ to the goldstino superfield, in the 
limit of vanishing momentum. We note that
one must also take $\Lambda\ra \infty$.

\noindent
{\bf b)} the validity and implications of the constraint  $X^2=0$ \cite{SK,Casalbuoni:1988xh}. 
This is supposed to decouple (project out)  the sgoldstino.  
In particular this can mean an infinite  sgoldstino mass \cite{Dudas:2011kt}, however 
 on dimensional grounds this is actually proportional to $f/\Lambda$. 
It is difficult to satisfy  a) and b) simultaneously in  general cases,
 because of opposite limits, of large $\Lambda$ and large sgoldstino mass, 
while  $f< \Lambda^2$. This situation is further complicated by the  
presence of more fields, some of which can  also contribute to (spontaneous) 
SUSY breaking.

The above problems were studied  for simple superpotentials, like
linear superpotentials  \cite{SK,Antoniadis:2011xi}
or with only one field breaking SUSY \cite{Dudas:2011kt}, with many
phenomenological applications  studied in \cite{pheno}.
 We investigate below problems a), b)  for  general $K$ and  $W$, 
 with more  fields contributing to SUSY breaking in the hidden sector. 
This helps a better understanding  of SUSY breaking and  its transmission 
to the visible sector via the coupling \cite{SK} of the  $X$ field to models like the 
Minimal Supersymmetric Standard Model (MSSM).

\section{(s)Goldstino and its relation to the  chiral superfield $X$.}

\subsection{Goldstino and sgoldstino eigenstates for arbitrary K, W.}

The starting point is the general Lagrangian 
\medskip
\bea\label{LKW}
L& =& \int d^4\theta\,\, K(\Phi^i,\Phi_j^\dagger)+
\Big\{\int d^2 \theta\,\, W(\Phi^i)+
\int d^2 \otheta\,\, W^\dagger(\Phi_i^\dagger)\Big\}
\nonumber\\
&=& \!\!\!
K_i^{\,\,\,j} \Big[\partial_\mu\phi^i\,\partial^\mu\phi_j^\dagger
+\frac{i}{2}\,\big( \psi^i\,\sigma^\mu\cD_\mu\opsi_j
-\cD_\mu \,\psi^i\sigma^\mu\opsi_j\big)
+ F^i\,F_j^\dagger\Big]
\nonumber\\[2pt]
&+&
\frac{1}{4}\,K_{ij}^{kl}\,\psi^i\psi^j\,\opsi_k\opsi_l
+\Big[\big(W_k - \frac{1}{2} \,K_k^{ij}\opsi_i\opsi_j\big) 
\,F^k - \frac{1}{2}\,W_{ij}\,\psi^i\psi^j
+h.c.\Big]
\label{generalKW}
\eea

\medskip\noindent
where we ignored a $(-1/4)\Box K$ in the rhs.
Here $K_i\equiv \partial K/\partial\phi^i$,
$K^n\equiv \partial K/\partial\phi^\dagger_n$,
$K_i^n\equiv \partial^2 K/(\partial\phi^i\,
\partial \phi_n^\dagger)$,
$W_j=\partial W/\partial \phi^j$, $W^j=(W_j)^\dagger$, etc,
with $W=W(\phi^i)$, $K=K(\phi^i,\phi_j^\dagger)$.

 Terms with
more than two derivatives of $K$ are suppressed by powers of $\Lambda$
which is the UV cutoff of the model, 
$K_{ij}^k\sim 1/\Lambda$, $K^{ij}_{km}\sim 1/\Lambda^2$, etc.
We also used the notation
\medskip
\bea
\cD_\mu \psi^l&\equiv&\partial_\mu\psi^l-\Gamma^l_{jk}\,(\partial_\mu\phi^j)\,\,\psi^k,
\qquad 
\Gamma^l_{jk}=(K^{-1})^l_m\,K_{jk}^m
 \nonumber\\[4pt]
 \cD_\mu \opsi_l&\equiv&\partial_\mu\opsi_l-\Gamma_l^{jk}\,
 (\partial_\mu\phi_j^\dagger)\,\,\opsi_k,
 \qquad
 \Gamma_l^{jk} =(K^{-1})^m_l\,K^{jk}_m
\eea

\medskip\noindent
Eq.(\ref{generalKW}) is the offshell form of the Lagrangian.
The eqs of motion for auxiliary fields 
\medskip
\bea
F_m^\dagger&=&-(K^{-1})^i_m \,W_i +({1}/{2})\,\Gamma_m^{lj}\,\opsi_l\opsi_j
\nonumber\\[4pt]
F^m&=& -(K^{-1})^m_i \,W^i +({1}/{2})\,\Gamma^m_{lj}\,\psi^l\psi^j 
\eea

\medskip\noindent
can be used to obtain the onshell form of $L$:
\medskip
\bea\label{ooss}
L&=&
K_i^{\,j}\,\Big[
\partial_\mu\phi^i\partial^\mu \phi_j^\dagger 
+
\frac{i}{2}\,\big(\psi^i\,\sigma^\mu\,\cD_\mu\opsi_j- \cD_\mu\psi^i\sigma^\mu\opsi_j\big)
\Big]
-W^k\,(K^{-1})^i_k\,W_i\nonumber\\[4pt]
&-&\frac{1}{2}\,\Big[\big(W_{ij}-\Gamma_{ij}^m\,W_m\big)\,\psi^i\psi^j+h.c.\Big]
+\frac{1}{4}\,R_{ij}^{kl}\,\psi^i\psi^j\,\opsi_k\opsi_l,
\qquad
R_{ij}^{kl}=K_{ij}^{kl}-K_{ij}^n\,\Gamma_n^{kl}
\eea

\medskip\noindent
Here $R_{ij}^{kl}$ is the curvature tensor and the potential of the model is
\medskip
\bea
V=W_i\, (K^{-1})^i_j\,W^j
\eea

\medskip\noindent
The derivatives of $K$, $W$ are scalar fields-dependent. 
In the following we always work in normal coordinates, in which case
$k^i_j=\delta_i^j$, $k^i_{jk...}=k_i^{jk....}=0$, 
where $k^{i...}_{j....}$ are the values of $K^{i....}_{j...}$
evaluated on the ground state  (denoted $\langle\phi^k\rangle$,
 $\langle F^k\rangle$, $\langle\psi^k\rangle\!=\!0$).
We denote the field fluctuations by $\delta\phi^i\!=\!\phi^i-\langle\phi^i\rangle$.
In normal coordinates $k_{ij}^{kl}$ used below is actually $k_{ij}^{kl}\!=\!R_{ij}^{kl}$.

From the eqs of motion for $F^i$, $\phi^i$, after taking the vev's, then
\medskip
\bea
k_i^j\,\langle F_j^\dagger\rangle 
+f_i
=0,\qquad
k_{im}^{\,j}\,\langle F^i\rangle \langle F_j^\dagger\rangle+f_{km}\langle F^m\rangle=0
\label{cc1}
\eea

\medskip\noindent
We denote by  $f_i$, $f_{ik}$, $f_{ijk}$, the values of corresponding, field dependent 
$W_i$, $W_{ik}$, $W_{ijk}$, evaluated on the ground state, so
\medskip
\bea\label{gs1}
f_i= W_i(\langle\phi^m\rangle), 
\quad f_{ij}=W_{ij}(\langle\phi^m\rangle), 
\quad f_{ijk}=W_{ijk}(\langle\phi^m\rangle), 
 \quad f^{i}=W^i(\langle\phi^m\rangle), \,\mbox{etc.}
\eea

\medskip\noindent
Eq.(\ref{cc1}) then becomes
\bea\label{sb}
\langle F_j^\dagger\rangle=-f_j,\qquad
f_{km}\langle F^m\rangle=0.
\eea

\medskip\noindent
To break supersymmetry, a non-vanishing vev of an auxiliary field is needed, 
which requires $\det f_{ij}=0$. 
The goldstino mass matrix is $(M_F)_{ij}=W_{ij}-\Gamma_{ij}^m\,W_m$ evaluated on the ground state,
giving $(M_F)_{ij}=f_{ij}$ in normal coordinates. 
A consequence of the last eq in (\ref{sb}) is that
the goldstino eigenvector (normalised to unity) is
\medskip
\bea\label{gg}
\tilde\psi^1= - \frac{\langle F_m^\dagger\rangle\,  \psi^m}{\big[
\langle F_i^\dagger\rangle \langle F^i\rangle\big]^{1/2}}
=\frac{f_m  \psi^m}{\big[f_i f^i \,\big]^{1/2}},\qquad m_{\tilde \psi^1}=0.
\eea

\medskip\noindent
Further, regarding the scalar sector, the mass matrix has the form
\medskip
\bea
M_b^2=
\left[\begin{array}{lr}
(V)^k_{\,\,l}    & (V)_{kl}\\
(V)^{kl}   & (V)^{\,\,\,l}_k
\end{array}
\right]
=
\left[
\begin{array}{lr}
f^{ik}\,f_{il} - k_{il}^{jk}\,\, f^i\,f_j     &   f_{jkl}\,f^j\\
f^{jkl}\,f_j           &   f^{il}\,f_{ik} - k_{ik}^{jl}\,\,f^i\,f_j
\end{array}\right],
\label{massb}
\eea

\medskip\noindent
where $V^k_{\,\,\,l}=\partial^2 V/(\partial\phi^l\partial \phi_k^\dagger)$, 
 $V_{kl}=\partial^2 V/(\partial\phi^l\partial \phi^k)$, 
etc, is evaluated on the ground 
state.

The two real components of the complex sgoldstino are mass degenerate only if in (\ref{massb})
the  off-diagonal (holomorphic or anti-holomorphic) blocks vanish.
For simplicity we assume  that this is indeed the case. 
This restricts the generality of our superpotential
by the condition  
\bea
f_{ijk}\,f^k=0,
\eea

\medskip\noindent
that we assume to be valid in this paper\footnote{
An  example when such condition is respected is for a  superpotential 
 of the type $W=f_1\Phi^1+\lambda/6\,(\Phi^2)^3$, with $\Phi^1$ 
breaking SUSY and $\Phi^2$ a matter field. This example will be considered later.}.
In this case, the block $(V)^k_{\,\,\,l}$  determines the mass
spectrum and eigenstates. The mass of (complex) sgoldstino obtained from this block
 must involve Kahler 
terms (their derivatives),
it cannot acquire  corrections from  $f_{ij}$ and it must be proportional to SUSY breaking, 
thus it depends on $f_i$, $f^i$. The only possibility in normal coordinates is to contract
the only non-trivial, non-vanishing tensor
 $k^{ij}_{kl}=R^{ij}_{kl}$  with $f_i$, $f^i$ and ensure the correct mass dimension and sign. 
The result for this (mass)$^2$ is given in the equation below and a discussion 
can  be found in  \cite{CS}. 
Finally since SUSY is broken spontaneously,
 the sgoldstino mass eigenvector is expected to 
have a form similar to that of goldstino  itself in eq.(\ref{gg}). 
Indeed, in the limit of ignoring the Kahler part of $(V)^k_{\,\,\,l}$
which is sub-dominant, of order $\cO(1/\Lambda^2)$, the sgoldstino 
is the (massless) eigenvector of $f^{ik}\,f_{il}$ matrix, and has the form:
\medskip
\bea\label{rq1}
\tilde\phi^1=
 \frac{f_m\,\delta \phi^m}{[f^i\,f_i]^{1/2}}+\cO(1/\Lambda^2),
\qquad 
m_{\tilde \phi^1}^2=-\frac{k^{ij}_{kl}\,f_i\,f_j\,f^k\,f^l}{f_m\,f^m}
\eea

\medskip\noindent
where $\delta\phi^m=\phi^m-\langle\phi^m\rangle$ is
the field fluctuation about the ground state. The mass of sgoldstino $\tilde\phi^1$
comes from D-terms,  which are $\cO(1/\Lambda^2)$.

Spontaneous SUSY breaking  suggests 
 the auxiliary of goldstino superfield should have a similar structure:
\medskip
\bea\label{aa2}
\tilde F^1 =\frac{f_m\, F^m}{[f^i\,f_i]^{1/2}}
\eea

\medskip\noindent
This is  verified onshell, when 
$F_i^\dagger=-W_i+\cO(1/\Lambda^2)$  is expanded about the ground state
to linear order fluctuations  $F_i^\dagger=-f_i-f_{im}\,\delta\phi^m+\cO(1/\Lambda^2)$. One
then finds from (\ref{aa2})\footnote{
Also at minimum,  $V$ should be just $\vert \langle\tilde F^1\rangle\vert^2$ 
 which is respected.}
\medskip
\bea\label{rq2}
\tilde F^1 =\frac{f_m\, (-f^m)}{[f^i\,f_i]^{1/2}}+\cO(1/\Lambda^2)
\eea

\medskip\noindent
For  illustration let us now consider in detail the case of only 
two fields present in Lagrangian (\ref{generalKW}),
and also present the expression of the second mass eigenvector. 
For the fermions, the mass eigenvectors (normalised to unity)
are given below, with $\tilde\psi^1$  the goldstino field:
\bigskip
\bea\label{13}
\tilde \psi^1&=&\frac{1}{[f_i\,f^i]^{1/2}}\,\Big[f_1 \,\psi^1+ f_2\,\psi^2\Big],
\qquad\qquad\qquad
m^2_{\tilde \psi^1}=0.
\nonumber\\
\tilde \psi^2&=&\frac{1}{[f_i\,f^i]^{1/2}}\,\Big[- (f_1/\rho) \,\psi^1+ f_2\,\rho \,\psi^2\Big], 
\qquad
m^2_{\tilde\psi^2}=f^{ij}\,f_{ij},\quad
\rho=\frac{\vert f_1\vert}{\vert f_2\vert}.
\eea

\bigskip\noindent
For the scalars sector, 
we find after some algebra the mass eigenstates\footnote{{
eqs.(\ref{13}) and (\ref{14}) are  multiplied in the rhs by $\vert f_2\vert/f_2$
which is set to unity  by phase rescaling $\Phi_2$.}} of $(M_b^2)=V^k_{\,\,\,l}$:
\bigskip
\bea\label{14}
\tilde \phi^1&=&\frac{1}{[f_i\,f^i]^{1/2}}\,\Big[
\,\,\,f_1 \,(1+\xi\,\tilde k_{11})\,\delta\phi^1
+ f_2\,(1+\xi\,\tilde k_{12})\, \delta \phi^2\Big],\nonumber\\
\tilde \phi^2&=&\frac{1}{[f_i\,f^i]^{1/2}}\,\Big[
- (f_1/\rho) \,(1+\xi\,\tilde k_{21})\,\delta \phi^1
+ f_2\,\rho \,(1+\xi\,\tilde k_{22})\,\delta \phi^2\Big],
\eea

\bigskip\noindent
with the notation
\bigskip
\bea\label{zz}
\tilde k_{11}\!\!\!&=&\!\!\! k^{ij}_{kl}\,\rho^k_i\,\,f_i f_j f^k f^l, \quad\quad
\tilde k_{12}= k^{ij}_{kl}\,\nu^k_i\,\,f_i f_j f^k f^l,\quad
\xi\equiv [ 2(f_k\,f^k)\,(f^{ij} f_{ij})]^{-1}\qquad \nonumber\\
\tilde k_{21}\!\!\!&=&\!\!\! k^{ij}_{kl}\,\sigma^k_i\,\,f_i f_j f^k f^l,\quad\quad
\tilde k_{22}= k^{ij}_{kl}\,\delta^k_i\,\,f_i f_j f^k f^l,\quad
\eea
where
\begin{align}
\rho^1_1&=2\,/\rho^2, \qquad
\rho^2_1=1+\rho^2+2/\rho^2,\qquad \quad
\rho^1_2=-3-\rho^2,     \qquad\qquad
\rho^2_2=-2,\nonumber\\
\nu^1_1&=-2,    \qquad\,\,\,\,\,
\nu^2_1=-(1-\rho^2),  \qquad\qquad\,\,\,\,
\nu^1_2=-(1-\rho^2),   \qquad\quad
\nu^2_2=2\,\rho^2,\nonumber\\
\sigma^1_1&=- 2,       \qquad\,\,\,\,\,
\sigma^2_1=1+1/\rho^2+2 \rho^2,\qquad\,\,\,
\sigma^1_2=-3-1/\rho^2,  \qquad\,\,\,\,
\sigma^2_2=2\,\rho^2,\nonumber\\
\delta^1_1&=2\,/\rho^2,     \qquad
\delta^2_1=-(1-1/\rho^2), \qquad\qquad
\delta^1_2=-(1-1/\rho^2), \qquad
\delta^2_2=-2.
\end{align}

\medskip\noindent
In our normal coordinates $k_{ij}^{kl}\!=\!R_{ij}^{kl}$.
We also find the masses:
\medskip
\bea\label{mass12}
m_{\tilde\phi^1}^2 = -\frac{k^{ij}_{kl}\,\,f_i f_j f^k f^l}{f_m f^m},
\qquad
m_{\tilde\phi^2}^2 = f^{ij} f_{ij}-k_{ik}^{jk}\,\,f^i f_j
+\frac{k^{ij}_{kl}\,\,f_i f_j f^k f^l}{f^m f_m}.
\eea

\medskip\noindent
We identify $\tilde\phi^1$ of (\ref{14}) as the sgoldstino,
 since its mass should not receive  corrections from $f_{ij}$,
in the limit of ignoring the curvature tensor corrections in (\ref{14}), and
 it has a form similar to that of goldstino eigenstate (\ref{gg}), (\ref{13}).
Regarding the auxiliary fields, one can show  that $\tilde F^2=O(1/\Lambda^2)$ 
and  that $\tilde F^1$ is that in  (\ref{rq2}).

We conclude that the goldstino superfield has the onshell SUSY form 
\bigskip
\bea\label{last}
\tilde\Phi^1\big\vert_{\rm on-shell} &=& 
\frac{f_k\,\delta \Phi^k}{[f^i f_i]^{1/2}}\big\vert_{\rm on-shell}\, +\cO(1/\Lambda^2)
\nonumber\\[5pt]
\delta\Phi^k\big\vert_{\rm on-shell} &\equiv 
&\delta\phi^k+\sqrt 2\,\theta\, \psi^k
+\theta\theta\,\,(-f^k).
\eea

\bigskip\noindent
where we used that $k^{ij}_{kl}=\cO(1/\Lambda^2)$  and that auxiliary 
fields are on-shell.

Eqs.(\ref{14}) to (\ref{last}) are valid under the assumption
that corrections suppressed by powers of $\Lambda$
are sub-leading to the superpotential SUSY corrections, proportional to 
$f_{ij}$, see also (\ref{massb}).
Let us introduce a parameter $\zeta$ equal to  the ratio  of the 
Kahler curvature tensor contracted by the  SUSY breaking scale(s) $f_i$ to 
the SUSY ``mass term'' ($f_{ij}$):
\medskip
\bea\label{param}
\zeta=\xi\,\tilde k_{ij}\sim 
\frac{k_{ij}^{kl}\,\,f^i f^j f_k f_l}{(f^p f_p)(f_{mn} f^{mn})}\sim 
\frac{m^2_{\mbox{\footnotesize sgoldstino}}}{f_{mn} f^{mn}}\leq 1.
\eea

\medskip\noindent
If $\zeta\leq 1$  the results of this section such as
(\ref{14}) and (\ref{last})  are valid and terms suppressed by high powers of $\Lambda$ 
can be neglected, as we actually did. For $\zeta\sim 1$ the eigenvectors have a more complicated
form (easily  obtained) and is  not presented here. The limit $\zeta\gg 1$ 
corresponds to  decoupling a massive sgoldstino
and  is discussed in Section~\ref{2.4}

We shall compare eq.(\ref{last}) to the chiral superfield $X$ that
breaks superconformal symmetry, conjectured in \cite{SK} to be equal, in the infrared
limit to the goldstino superfield  $\tilde\Phi^1$.

\subsection{The chiral superfield X and  its low-energy limit.}

Let us explore the properties of the superconformal symmetry breaking
chiral superfield $X$ and examine its relation to the goldstino
superfield found earlier. The definition of $X$ is 
\medskip
\bea
\overline D^{\dot \alpha} \cJ_{\alpha\dot\alpha}
=D_\alpha X, 
 \qquad\quad X\equiv(\phi_X,\psi_X,F_X)
\eea

\medskip\noindent
where $\cJ$ is the Ferrara-Zumino current \cite{Ferrara}. 
For a review of this topic, see for example section 2.1 in \cite{SK}.
$\psi_X$ is related to the supersymmetry current and $F_X$ to the energy-momentum tensor.
For the general, non-normalizable action  in (\ref{LKW}), 
this equation has a solution \cite{Clark:1995bg}
\medskip
\bea\label{eq9}
X=4\,W- \frac{1}{3} \,{\overline D}^2\,K - \frac{1}{2} \,{\overline D}^2 
Y^\dagger(\Phi^\dagger)
\eea

\medskip\noindent
We find the component fields of $X$  to be
(ignoring the improvement term $\overline D^2 Y^\dagger(\Phi^\dagger)$):
\medskip
\bea
\label{compX}
\phi_X&=& 4\,W(\phi^i)+\frac{4}{3}\,\Big[K^j\,F_j^\dagger
 -\frac{1}{2} K^{ij}\,\opsi_i\opsi_j\Big]
\nonumber\\[4pt]
\psi_X&=& 
\psi^k\,\frac{\partial \phi_X}{\partial \phi^k}
-\frac{4\,i}{3}\sigma^\mu\,\big( K^j\,\partial_\mu\opsi_j
+K^{ij}\,\opsi_j\partial_\mu\phi_i^\dagger\big)
\nonumber\\[3pt]
F_X\! & = &\!\!\!
 F^i\,\frac{\partial \phi_X}{\partial \phi^i}\! -\!
\frac{1}{2} \psi^i\psi^j
\frac{\partial^2 \phi_X}{\partial \phi^i\partial\phi^j}
\!+\!\frac{4}{3}\,\Big\{
K_i^j\,\Big[\partial_\mu\phi^i\partial^\mu \phi_j^\dagger
+\frac{i}{2}\,\big(\psi^i\sigma^\mu D_\mu\opsi_j 
\!- \! D_\mu \psi^i\,\sigma^\mu \opsi_j\big)\Big]
\nonumber\\
&-&
\partial_\mu
\Big(
K^j\,\partial_\mu \phi_j^\dagger
-\frac{i}{2} K_i^j\psi^i\sigma^\mu\opsi_j\Big)
\Big\}
\eea

\medskip\noindent
In these relations all derivatives are scalar-fields dependent quantities.
As a side-remark, one
 also notices that the integer powers $n\geq 1$ of these components  have a nice
compact structure:
\medskip
\bea\label{powers}
\phi_{X^n}\!\!&=& (\phi_X)^n,\qquad\qquad  (n\geq 1)\nonumber\\[5pt]
\psi_{X^n}\!\!&=& n \,(\phi_X)^{n-1} \,\psi_X
 =
 \psi^j\,\frac{\partial \phi_{X^n}}{\partial\phi^j}+\cO(\partial_\mu)
\nonumber\\
F_{X^n}\!\! &=&\!\!
 n\,(\phi_X)^{n-2}\,\Big[\phi_X\,F_X-\frac{n-1}{2}\,\,\psi_X\psi_X\Big]
=\!
F^j\,\frac{\partial \phi_{X^n}}{\partial\phi^j}-\frac{1}{2}\psi^i\psi^j\,
\frac{\partial^2\phi_{X^n}}{\partial\phi^i\partial\phi^j}+\cO(\partial_\mu),\,\,\,
\eea

\medskip\noindent
where the terms $\cO(\partial_\mu)$  vanish in the infrared limit of zero momenta.

Notice that  in the  leading (zero-th) order in $1/\Lambda$, the only dependence of these 
components on the Kahler comes through $\phi_X$ via its term  
$K^j\,F_j^\dagger$, with additional contributions, fermionic dependent being\footnote{
The bracket in $F_{X^n}$ is SUSY invariant 
for $n=2$, and then  $F_{X^2}=0$ is invariant.} $\cO(1/\Lambda)$.

From (\ref{compX}) we expand $X$ about the ground state
and  denote $w=W(\langle\phi^k\rangle)$.
Keeping linear fluctuations in fields, one obtains from eq.(\ref{compX})  that
\medskip
\bea\label{onshellx}
\phi_X &=& 4\,w+\frac{8}{3}\,f_j\,\delta\phi^j+\cO(1/\Lambda),\nonumber\\
\psi_X &=& \frac{8}{3}\,f_k\,\psi^k +\cO(1/\Lambda),
\nonumber\\
F_X &=& \frac{8}{3}\,f_k\,(-f^k) 
- 4 \,f^k f_{km}\,\delta\phi^m-\frac{4}{3}\,f^k\,\delta F_k^\dagger
+ \frac{8}{3} f_k\,\delta F^k
+\cO(1/\Lambda).
\eea

\medskip\noindent
Up to a constant we can  write,
using eq.(\ref{last}), that onshell-SUSY:
\medskip
\bea
X\big\vert_{\rm on-shell}\!
&=&
\frac{8}{3}\,f_k\,\big[\delta\phi^k+\sqrt 2\, \theta\, \psi^k+\theta\theta\,(-f^k)\big]
+\cO(1/\Lambda)
\nonumber\\
&=&\frac{8}{3}\,
f_k\,\delta\Phi^k\big\vert_{\rm on-shell}+\cO(1/\Lambda)
\eea

\medskip\noindent
Comparing this result  against that for the goldstino superfield of (\ref{last}),
one has
\medskip
\bea\label{relation}
X\big\vert_{\rm on-shell}=
\frac{8}{3}\,\sqrt{f_i f^i}\,\tilde\Phi^1\big\vert_{\rm on-shell}+\cO(1/\Lambda).
\eea

\medskip\noindent
Note that the $X$ field goes to the (onshell) goldstino field in the limit
of vanishing momentum and in addition  $\Lambda\ra \infty$ when higher dimensional
terms in the Kahler potential decouple.
This  clarifies the relation 
 between the goldstino and the superconformal symmetry breaking superfields
for general $K$ and $W$, in the presence of  more sources of SUSY breaking,
and  is one of the results of this work.
All directions of supersymmetry breaking contribute to the relation 
between these two superfields.

\subsection{Further properties of the field $X$.}

Let us compute the onshell form of $X$ by  eliminating
 the auxiliary fields $F^k$  in (\ref{compX}) and then examine
 under what conditions $X^2$ could vanish.  The results below are
valid up to $\cO(\partial_\mu)$, where all  terms $K^j$, $W_k$... etc  
are actually scalar-fields dependent.  One has
\medskip
\bea\label{ons1}
\phi_X&=& \sigma+\sigma^{mn}\,\,\opsi_m\opsi_n,
\nonumber\\[4pt]
\psi_X& =& \frac{8}{3}\,\psi^k\,W_k,
\nonumber\\[4pt]
F_X&=& \beta+\beta_{mn}\,\psi^m\psi^n+
\beta_{mn}^{kl}\,\,(\psi^m\psi^n)\,\,(\opsi_k\opsi_l),\qquad\qquad\qquad\qquad
\eea
\medskip\noindent
where
\bea
\sigma&=&4\,W-(4/3)\,\, K^l\,(K^{-1})^k_l\,\,W_k, 
\nonumber\\[6pt]
\sigma^{mn} &=& (2/3)\,\big[K^l\,\Gamma_l^{mn}
-K^{mn}\big]
=\cO(1/\Lambda),
\nonumber\\[6pt]
\beta&=&-(8/3)\,\,W_m\,(K^{-1})^m_k\,\,W^k,
\nonumber\\[6pt]
\beta_{mn}& = & 2\,\big[\,W_{k}\,\Gamma^k_{mn}-W_{mn}\big]
=-2 \,W_{mn}+ \cO(1/\Lambda),
\nonumber\\[6pt]
\beta_{mn}^{kl}&=&
(1/3)\,\big(K_{mn}^{kl}-K_{mn}^j\,\Gamma_i^{kl}\big)\equiv
(1/3)\, R_{mn}^{kl}
=\cO(1/\Lambda^2).\qquad\quad
\eea

\medskip\noindent
Here  we made explicit the terms which are suppressed by powers of the cutoff scale.

In \cite{SK,Casalbuoni:1988xh}  it was used that the constraint $X^2=0$
projects out the sgoldstino field\footnote{Additional constraints for 
matter superfields can be used to decouple superpartners at low energy.}.
In a strict sense this constraint is  valid only in the limit of an infinite 
sgoldstino mass. So the problem is that one has
an expansion in $1/\Lambda$  of the initial Lagrangian
which can conflict with an  expansion in the  inverse sgoldstino mass,
$1/m^2_{\tilde\phi^1}\sim 1/(f_i^2/\Lambda^2)=\Lambda^2/f_i^2$,
that decouples the sgoldstino.
The effective Kahler terms must give a mass to sgoldstino (which would 
otherwise be massless  at tree level in spontaneous Susy breaking),
and must simultaneously be large enough for the sgoldstino to 
decouple at low energy.  The two expansions may  have only a very small overlap
region of simultaneous convergence. 

To have $X^2=0$ it is necessary and sufficient to have $F_{X^2}=0$. 
This can be checked directly. If for example $F_{X^2}=0$ then one immediately shows that
 $\phi_X \sim \psi_X\psi_X$ so $X^2=0$. Let us compute the value of $F_{X^2}$ in general,
using (\ref{ons1}). One has
\bigskip
\bea\label{yy}
F_{X^2}\!\! &=&\! \alpha
\!+\!
\alpha_{mn}\,(\psi^m\psi^n)
\!+\!
\lambda^{mn}\,(\opsi_m\opsi_n)
\!+\!\nu_{mn}^{kl}\,(\psi^m\psi^n)(\opsi_k\opsi_l)
\!+\!
\xi_{mn}\, (\psi^m\psi^n)(\opsi_1\opsi_1)(\opsi_2\opsi_2)
\nonumber\\[7pt]
&=&
 \alpha
+
\alpha_{mn}\,(\psi^m\psi^n)+\cO(1/\Lambda)
\eea
with
\bea\label{omegas}
 \alpha &=& 2\sigma\beta,
\qquad\qquad\qquad
\alpha_{mn}= 2\sigma\beta_{mn}-\frac{64}{9} W_m W_n,
\eea

\medskip\noindent
while the remaining coefficients are suppressed, of order $\cO(1/\Lambda)$
 or higher\footnote{The expressions of these terms are
$ \lambda^{mn}= 2\,\beta\,\sigma^{mn}=\cO(1/\Lambda)$, 
$\nu_{mn}^{kl}= 2\,\big[\sigma\,\beta_{mn}^{kl}+\beta_{mn}\sigma^{kl}\big]=\cO(1/\Lambda)$,
$\xi_{mn}=
2\,\big[\sigma^{11}\,\beta_{mn}^{22}+\sigma^{22}\,\beta_{mn}^{11}-2\sigma^{12}\,\beta^{12}_{mn}
=\cO(1/\Lambda)
\big]$.} and vanish at large $\Lambda$. In this limit only,
expanding (\ref{yy}) about the ground state
(or using (\ref{onshellx})) we find 
\medskip
\bea
F_{X^2} 
=
-\frac{64}{9}\,
(2\,f_k f^k f_l\,\,\delta\phi^l+f_k f_l\,\,\psi^k\psi^l)
+\cO(1/\Lambda)
\label{fx2}\eea

\medskip\noindent
up to a constant ($\propto w$). To see if  $F_{X^2}$ and thus $X^2$ vanish
(hereafter this is considered up to $\cO(1/\Lambda)$ terms)
 after decoupling massive scalar fields,
one should integrate out $\delta\phi^k$, $k=1,2...$,   via the eqs of motion.
With  $\delta\phi^k$ expressed in terms of the light fermionic and other 
scalar degrees of freedom, one checks in this way if $X^2=0$, without the need of 
computing the mass eigenstates\footnote{For one field breaking SUSY, 
$f^1\not=0$, ($f^{j\not=1}\!=\!0$) 
 and if $w=0$,   $F_{X^2}=0$  if $\delta\phi^1=\psi^1\psi^1/(-2\,f^1)$ see \cite{SK}.}.
In general,  upon  integrating out the sgoldstino and additional massive scalars,  
$X^2$ necessarily contains  terms suppressed by the sgoldstino mass.
In most  cases $X^2$  does not vanish anymore if this mass is {\it finite},  except in 
specific cases,  due to additional simplifying assumptions (symmetries, etc) for the terms 
(e.g. $k^{ij}_{mn}$) of  the Lagrangian.  In these cases the convergence problem mentioned 
earlier is not an issue. We  discuss such a case in the next two sections.

\subsection{Decoupling all scalar fields, for  vanishing SUSY mass terms.}\label{2.4}

Let us consider the special case of a vanishing SUSY term, i.e. 
$f_{ij}=0$ or assume it is much smaller than the Kahler terms 
in the mass matrix  of eq.(\ref{massb}). 
We consider only two  fields present in the  Lagrangian (\ref{LKW}), 
both of which can contribute to the  SUSY breaking. This can  be
generalised to more fields. For $f_{ij}=0$  we have two massless fermions.
As a result, both  scalar fields,  which
are massive (via Kahler terms), can be integrated out and expressed 
in terms of these massless fermions, without special restrictions for
scales present.
We shall do this and then examine under what conditions $X^2$ vanishes
after decoupling.

Regarding the relation of $X$ to the goldstino superfield, in this case
it is difficult to define the latter, as both fields contribute to SUSY breaking
and they can also mix.
In previous sections, see eq.(\ref{param}), (\ref{relation}) 
the superpotential ($f_{ij}$) terms were dominant  and
Kahler terms were a small correction ($\propto 1/\Lambda^2$)
to the scalars mass matrix (\ref{massb}),  while here the situation is reversed. 
Eq.(\ref{massb}) with vanishing 
off-diagonal blocks and vanishing  $f^{ik}\,f_{il}$ gives a mass matrix 
$M_b^2\!=\!(V)_l^k\!=\!-k_{il}^{jk}\,f^i f_j$ in basis $\delta\phi^{1,2}$
 with eigenvalues 
\medskip
\bea\label{pp}
m_{\tilde \phi^{1,2}}^2\!=\!\frac{1}{2}\Big[-k_{mk}^{mj}\,f^k f_j\pm\sqrt \Delta \Big]+\cO(f^{ij} f_{ij}),
\quad
\Delta\!=\! (k_{mk}^{mj}\,f^k f_j)^2-4\,\det( k_{mk}^{pj}\,f^k f_j)
\eea

\bigskip\noindent
where all indices take values 1 and 2, and the determinant is over the free indices.
This is the counterpart to the result in  (\ref{mass12}).
The eigenvectors can also be found\footnote{They are
$\tilde\phi^{1,2}\!=\!(1/\vert \tilde\phi^{1,2}\vert)\{(2 k_{1 j}^{2k} f_k f^j)^{-1}
\big[ (k_{1m}^{1 n}-k_{2m}^{2n}) f^m f_n \!\pm\! \sqrt \Delta\big]\delta\phi^1
\!+\!\delta \phi^2\}+\cO(f^{ij}f_{ij})$ ($\vert\tilde\phi^{1,2}\vert$: norm).}.
How do we identify the sgoldstino? The term $f_{ij} f^{jk}$ which defined
in (\ref{massb}) the leading contribution to the
mass matrix and eigenvectors, is vanishing, so it cannot be used.
One can identify  the sgoldstino from a transformation that ensures 
that only  one linear combination of auxiliary fields
breaks supersymmetry. The scalar in the same supermultiplet 
 is then the sgoldstino; further, if no mixing is induced by Kahler 
curvature terms (this mixing is controlled by $k^{ij}_{mn}$  and is therefore
UV and model-dependent) then this state is also a mass eigenstate.
To this end  define  new superfields
\medskip
\bea\label{re}
\tilde\Phi^{1}&=&\frac{1}{[ f_k f^k]^{1/2}}\,(f_1\,\delta
\Phi^1+f_2\,\delta \Phi^2),
\nonumber\\
\tilde\Phi^{2}&=&\frac{1}{[ f_k f^k]^{1/2}}\,(-(f_1/\rho)\,\delta \Phi^1+f_2\,\rho\,\delta\Phi^2),
\qquad\rho=\vert f_1\vert/\vert f_2\vert.
\eea

\medskip\noindent
where $\delta\Phi^j\!=(\delta\phi^j,\psi^j,F^j)$.
$\tilde\Phi^1$ is inferred from the auxiliary fields combination
and $\tilde\Phi^2$ was determined by  unitarity arguments.
One can apply this transformation to the original Lagrangian, then if
 scalar components are not mixing,
$\tilde\phi^1$ is also a mass eigenstate\footnote{
This requires  a diagonal mass matrix in $\delta\phi^{1,2}$ initial basis, i.e.  
$k_{2m}^{1k}\,f_k f^m=k_{1m}^{2k}\,f_k f^m=0$  (for a vanishing $f_2$ this means
$k^{11}_{12}=k^{12}_{11}=0$).  In this case the masses are
$m_{\tilde\phi^1}^2=-k^{1m}_{1n}\,f_m f^n$, $m_{\tilde\phi^2}^2=-k^{2m}_{2n}\,f_m f^n$.}.

Let us now discuss the decoupling of the scalars  and check
under what conditions $X^2$  can vanish, 
 without demanding an  infinite sgoldstino mass (which
would bring  convergence problems). We integrate the scalars, 
so in the low energy they are combinations of the light/massless fermions.
To this purpose, we do not need to identify the sgoldstino.
From  (\ref{ooss}), the eq of motion of scalar field 
$\phi^\dagger_l$, at zero-momentum, is
\medskip
\bea
W^{kl} (K^{-1})^i_k W_i
+
W^k (K^{-1})^{il}_k W_i
+\frac{1}{2} \big(W^{ijl}-\partial^l (\Gamma_m^{ij}\,W^m)) \opsi_i \opsi_j
-
\frac{1}{2}\,\partial^l\Gamma_{ij}^m\,W_m\psi^i\psi^j\!=\!0.
\eea

\medskip\noindent
We expand this  about the ground state, in normal coordinates and
use our simplifying assumptions 
\bea\label{ccon}
f_{ij}=0,\qquad  f^{ijl} f_l=0, 
\qquad
\mbox{and}\qquad  
f^{ijlm}=0.
\eea

\medskip\noindent
The result is 
\medskip
\bea\label{eqm}
k_{kj}^{il}\,\delta\phi^j\,\,f^k f_i+\frac{1}{2}\,k_{ij}^{lm}\,f_m\,\psi^i\psi^j
-\frac{1}{2}\,f^{ijl}\,\opsi_i\opsi_j+\cO(1/\Lambda^3)=0,\qquad i,j,k,l,m=1,2.
\eea

\medskip\noindent
Taking $l=1, 2$, we solve this system for $\delta\phi^{1,2}$ to find
\medskip
\bea\label{ss22}
\delta\phi^1&=& \frac{1}{2 \det(k_{lm}^{kn}\,f_n f^m)}\,
\Big[
A_{ij}\,\,\psi^i\psi^j+ B^{ij} \,\,\opsi_i\opsi_j\Big]+\cO(1/\Lambda)
\nonumber\\
\delta\phi^2&=& \frac{1}{2\det(k_{lm}^{kn}\,f_n f^m)}\,
\Big[
 C_{ij}\,\,\psi^i\psi^j+ D^{ij} \,\,\opsi_i\opsi_j)\Big]+\cO(1/\Lambda)
\eea

\medskip\noindent
with
\medskip
\bea
A_{ij}&=&  \big(k_{ij}^{2p}\,k^{1 r}_{2s}-k_{ij}^{1p}\,k^{2r}_{2s}\big)\,f^r f_s f_p,
\quad
B^{ij}= -  f^{ij2}\,k_{2s}^{mr}\,f^s f_m f_r  (f_1)^{-1},
\nonumber\\
C_{ij}&=& \big(k_{ij}^{1p}\,k^{2r}_{1s}-k_{ij}^{2p}\,k^{1r}_{1s}\big)\,f^r f_s f_p,
\quad
D^{ij}=- f^{ij1}\,k_{1s}^{mr}\,f^s f_m f_r (f_2)^{-1}.
\eea

\medskip\noindent
The fields are suppressed by the mass of sgoldstino since 
the determinant in the  denominator
is a product of scalar masses, see (\ref{pp}); but due to interactions ($f^{ijl}$), by
counting the mass dimensions, the expansion can also be regarded as proportional 
to $\Lambda^2/f_i^2$ ! Indeed:
\medskip
\bea
\delta\phi^1\propto
\frac{1}{m^2_{\mbox{\footnotesize sgoldstino}}}\,\Big[A_{ij} \,\psi^i\psi^j
+ B_{ij}\,\opsi_i\opsi_j\Big], \qquad\mbox{\rm (similar for $\delta\phi^2$).}
\eea

\medskip\noindent
This is the sgoldstino decoupling limit, opposite to that considered in eq.(\ref{param}).

To  check if  $X^2$ vanishes for finite sgoldstino/scalars masses,
we use the result of (\ref{fx2}). From (\ref{ss22}) one finds that this happens if
\medskip
\bea\label{tr}
(f_k f^k)\,\big[f_1\,A_{ij}+f_2\,C_{ij}\big]+\det(\tilde k_{n}^{m})\,f_i f_j =0,\qquad 
f_1\,B^{ij}+f_2\,D^{ij}=0
\eea

\medskip\noindent
with  $\tilde k^m_n\equiv k^{mr}_{ns}\,f_r f^s$. The last two equations can be re-written as
\medskip
\bea\label{dec}
f_p\,\big[
k_{ij}^{2p}\,(f_1\,\tilde k^1_2 - f_2 \,\tilde k^1_1)-
k_{ij}^{1p}\,(f_1\,\tilde k^2_2 - f_2\, \tilde k_1^2)
\big]
+ \det (\tilde k_{n}^{m})\,\frac{f_i f_j}{f_k f^k}
&=& 0
\nonumber\\[6pt]
k_{ls}^{mr}\,f^{ijl}\,f^s f_m f_r&=& 0,
\eea

\medskip\noindent
where  $i,j$ are fixed to any value, 1,2.
If these relations are respected one has in the model $X^2=0$ for a finite sgoldstino mass
and trilinear interactions in the superpotential.
These relations ultimately imply some constraints  for the curvature tensor 
and thus for the UV regime. The first relation in (\ref{dec}) simplifies 
further in specific cases, for example if $\tilde\phi^1$ of (\ref{re}) 
is also a mass eigenstate which happens for $\tilde k^1_2=\tilde k^2_1=0$.
Conditions (\ref{dec}) can be generalised to more fields and should be verified in  
those applications in which the constraint $X^2=0$ was used.
These conditions would also be recovered with our definition of the goldstino superfield
in (\ref{re}). With this definition, the above relations are obtained 
by demanding (onshell) $(\tilde\Phi^1)^2=0$, or equivalently
 $F_{(\tilde\Phi^1)^2}=2 \tilde F^1\,\tilde\phi^1-\tilde\psi^1\tilde\psi^1=0$.

To illustrate some implications, let us take in eq.(\ref{ss22}) the limit of
only one field breaking supersymmetry, i.e. assume $f_2=0$. One finds
\bigskip
\bea\label{st}
\delta\phi^1&=&-\frac{\psi^1\psi^1}{2\,f^1}+\frac{\det(k_{2j}^{1i})}{\det (k^{1m}_{1n})}
\,\frac{\psi^2\psi^2}{2\,f^1} 
- \frac{ \,k_{21}^{11}\,f^{ij2}}{\det (k^{1m}_{1n})}\,\frac{\opsi_i\opsi_j}{2\,\vert f_1\vert^2}
+\cO(1/\Lambda)
\nonumber\\
\delta\phi^2&=&-\frac{\psi^1\psi^2}{f^1}
+\frac{k_{11}^{12}\,k_{22}^{11}-k_{11}^{11}\,k_{22}^{12}}{\det (k^{1m}_{1n})}
\,\frac{\psi^2\psi^2}{2\,f^1} 
+ \frac{ \,k_{11}^{11}\,f^{ij2}
}{\det (k^{1m}_{1n})}\,\frac{\opsi_i\opsi_j}{2\,\vert f_1\vert^2}
+\cO(1/\Lambda)
\eea

\bigskip\noindent
This is the general result for the scalars as functions of the massless fermionic 
fields,  when  superpotential interactions are present\footnote{As usual, in the normal coordinates 
used here $k_{ij}^{mn}=R_{ij}^{mn}$.}.
This result recovers eqs.(33), (37) in \cite{Antoniadis:2011xi}  
but have   additional 
corrections due to superpotential couplings.
The terms proportional to $f^{ij2}$ in both
$\delta\phi^{1,2}$ are actually dominant, since they grow like $\Lambda^2$,
as it can be seen from the mass dimensions of the $k^{ij}_{lm}$.
The other terms, coefficients of $\psi^1\psi^1$ and $\psi^2\psi^2$
are actually independent of $\Lambda$, although for $\psi^2\psi^2$ they 
involve UV details\footnote{Similar effects were discussed
  in \cite{Dudas:2011kt,Antoniadis:2011xi}.}.

From (\ref{fx2}) one obtains  $F_{X^2}\propto (2\,f^1\,\delta\phi^1+\psi^1\psi^1)$  
\cite{SK}, which we demand to vanish.
 Using $\delta\phi^1$ of (\ref{st}) or directly from the two equations in (\ref{dec}),
one finds that  $X^2=0$ if
\medskip
\bea
\det(k^{1i}_{2j})=0,\qquad \mbox{and}\qquad f^{ij2}\,k^{11}_{12}=0.
\label{tt}
\eea

\medskip\noindent
These constraints are a particular case of the  general conditions in (\ref{dec}).
If the Lagrangian  respects these conditions, one can
 have\footnote{up to $\cO(1/\Lambda)$ corrections.}
$X^2=0$ in the presence of trilinear interactions, with one field breaking SUSY
and finite mass sgoldstino. Finally, let us add that eq.(\ref{st}) and 
conditions (\ref{tt}) simplify further  if one demands $\tilde\phi^1$ of (\ref{re})
be also a mass eigenstate which only happens under a special, additional UV assumption: 
$k^{11}_{12}\!=\! k^{12}_{11}\!=\!0$. Then condition (\ref{tt})  reduces to $k^{11}_{22}\!=\!0$. 
This is however a particular case, not considered further.

\subsection{Decoupling the sgoldstino in the presence of a light matter field.} 

There are situations when the sgoldstino is significantly heavier than
other scalar (matter) fields and is the first or the sole field
to decouple at low energy. If so, under what conditions is $X^2=0$?
 To examine this briefly,
consider the case of the previous section, of two fields $\Phi^{1,2}$
in the Lagrangian, with  a simple superpotential
\medskip
\bea
W=f_1\Phi^1+\frac{\lambda}{3!}\,(\Phi^2)^3,
\label{ll}
\eea

\medskip\noindent
 So $\Phi^1$ breaks supersymmetry and we also assume that  
its scalar component (sgoldstino) is  much heavier  than the second scalar (matter) field 
belonging to $\Phi^2$. One can ensure such mass hierarchy  by assuming that
$\det (k_{1m}^{1n})$ is small enough, see (\ref{pp}).
Although we do not consider here
the extreme case when it actually vanishes, in that case one 
has  (if $k_{11}^{11}+k_{12}^{12}<0$) that
$h_{11}^{11} \,k_{12}^{12}- k_{12}^{11}\,k^{12}_{11}\approx 0, \, 
m_{\phi^1}^2\approx - (k_{11}^{11}+k_{12}^{12})\,f_1^2,\,
m_{\phi^2}^2\approx 0$.
Let us then integrate out the sgoldstino.
Its eq of motion, from Lagrangian (\ref{ooss}) or (\ref{eqm}), is
\medskip
\bea\label{solphi1}
\delta\phi^1&=&
-\,\frac{1}{f_1\,k_{11}^{11}}\,\,\,\Big[(1/2)\,k_{mn}^{11}\,\psi^m\psi^n+\,f_1\,\delta\phi^2\,k_{12}^{11}
\Big]+\cO(1/\Lambda),
\eea

\medskip\noindent
which is a function of the light scalar and  massless fermions.
Using   (\ref{solphi1})   one finds
\medskip
\bea
F_{X^2}\!&=&
\!\!\frac{64}{9}\,\frac{(f_1)^2}{k_{11}^{11}}\,
\Big[
2\,k_{12}^{11}\,(f_1\,\delta\phi^2+\psi^1\psi^2)+
k^{11}_{22}\,\psi^2\psi^2
\Big]
+\cO(1/\Lambda)
\label{xp}
\eea

\medskip\noindent
For any value of the scalar matter field ($\phi^2$), 
with sgoldstino decoupled at  finite mass,
one can thus have $F_{X^2}=X^2=0$ only if $k^{11}_{12}=0$, $k^{11}_{22}=0$ and for a large $\Lambda$.
These conditions can be compared to those when both scalars are decoupled shown in
(\ref{tt}) (with $f^{ij2}\ra\lambda$).
Therefore the action  for which the formalism of \cite{SK} applies with $X^2=0$, 
has $K$ given by
\medskip
\bea
K&=&\Phi_1^\dagger\Phi_1+\Phi_1^\dagger\Phi_1
+k_{11}^{11}\,(\Phi_1^\dagger\Phi_1)^2
+
k_{22}^{22}\,(\Phi_2^\dagger\Phi_2)^2
+
\big[ \,k_{22}^{21}\,(\Phi_2^\dagger\Phi_2)(\Phi_2\Phi_1^\dagger)+h.c.\big]\nonumber\\[3pt]
&+&
k_{12}^{12}\,(\Phi_1^\dagger\Phi_1)(\Phi_2^\dagger\Phi_2)
+\cO(1/\Lambda^3)
\eea

\medskip\noindent
with a nontrivial superpotential as in  (\ref{ll}) and a  finite sgoldstino mass.

In the Lagrangian obtained after decoupling $\delta\phi^1$ 
one can now also integrate out $\delta\phi^2$ and obtain a solution for it
 as in  (\ref{st}) but with the replacement
$f^{ij2} \opsi_i\opsi_j\ra\lambda\opsi_2\opsi_2$. This solution, if used in (\ref{solphi1}), 
brings  $\delta\phi^1$ to the form shown in (\ref{st}), as expected. With 
this  $\delta\phi^{2}$  one then  easily verifies   that $F_{X^2}$ of  (\ref{xp}) becomes
\medskip
\bea\label{fxsq}
F_{X^2}=\frac{64}{9}\,\frac{-(f_1)^2}{\det (k^{1m}_{1n})} 
\Big\{
\det(k_{2j}^{1i})\,\psi^2\psi^2 -\frac{\lambda\,k^{11}_{12}}{f_1}\,\,\opsi_2\opsi_2\Big\}
\eea
and
\bea
X^2=-\frac{F_{X^2}}{f^1}\,\,
\Big\{
\frac{-1}{2\,f^1}\,\Big(\psi^1\psi^1 +\frac{9}{128}\,\frac{F_{X^2}}{(f_1)^2}\Big)
+\sqrt 2 \,\theta\,\psi^1+\theta\theta\,(-f^1)\Big\}+\cO(1/\Lambda).
\eea 

\medskip\noindent
Therefore $F_{X^2}$ vanishes and  so does $X^2$ provided that
 $\det(k_{2m}^{1l})=0$ and $\lambda \, k^{11}_{12}=0$, and this recovers the 
result in eq.(\ref{tt}) when both scalars are decoupled.

Higher powers of $X$ can  vanish with weaker restrictions. This is actually
 expected from the properties
of the Grassmann variables. Indeed,
one shows that in onshell-SUSY case after decoupling only the sgoldstino 
($\delta\phi^1$) then:
\medskip
\bea
{X^3}\propto \frac{k_{12}^{11}}{k_{11}^{11}}\,
f_1\times \mbox{(function of $\delta\phi^2$, $\psi^{1,2}$)}.
\eea

\medskip\noindent
This vanishes for any $\delta\phi^2$ and finite sgoldstino mass, provided that
$k_{12}^{11}=0$ which is a weaker constraint than that found for $X^2$.
Higher powers of $X$ show that $k_{12}^{11}=0$ is still needed 
 for $X^4$ to vanish for any light matter field,  because in (\ref{solphi1})
 $\delta\phi^2$  is multiplied by $k_{12}^{11}$. Recall however that $k^{11}_{12}$
vanishes if there is no scalars mixing induced by Kahler curvature terms
i.e. if $\phi^1$ of (\ref{ll}) is also a mass eigenstate.

\section{Conclusions.}

In this work we considered the relation of the superconformal
 symmetry breaking chiral superfield $X$
and the goldstino superfield, in effective models with low scale of SUSY breaking,
when transverse gravitino couplings are negligible relative to
their longitudinal counterparts of its goldstino component.
 The models considered have a general Kahler ($K$)  and superpotential ($W$) 
with more sources of  supersymmetry breaking.

 In this case we verified the 
conjecture that the  superfield $X$ becomes  the goldstino superfield
 in the limit of zero-momentum and, in addition, 
 $\Lambda\ra \infty$, where $\Lambda$ is the UV cutoff. This happens
when the higher dimensional Kahler terms are sub-leading to the supersymmetric mass terms
in the scalar mass matrix. 
For vanishing SUSY mass terms, but otherwise rather general $K$ and $W$ 
we also investigated the decoupling of the massive scalars simultaneously or separately.
In this case we identified the  conditions  for which the 
sgoldstino  decoupling condition $X^2=0$
is still satisfied in the presence of additional fields,
  for a  {\it finite} sgoldstino mass. This is important to
ensure  that the effective expansion ($\propto 1/\Lambda$) of the Lagrangian does not 
conflict with the sgoldstino decoupling limit (of small 
$\propto 1/m^2_{\mbox{\footnotesize sgoldstino}}\sim \Lambda^2/f_i^2$ where  $f_i$ is the SUSY
 breaking scale). The above conditions are lifted 
 in the  {\it formal} limit of very large sgoldstino mass (or when
all scalar and fermion fields  other than the Goldstino fermion have all
non-zero masses and are integrated out);  then, in the far infrared
(i.e. far below any of these mass scales and at zero momentum) one recovers 
the relation $X^2=0$ of the Akulov-Volkov action for the goldstino.

One can reverse the above arguments and conclude that the use of the 
constraint   $X^2=0$, although appealing and apparently UV independent, is of somewhat 
restricted applicability  in the case of general $K$, $W$ (with massless fields present,
 additional SUSY breaking fields and interactions, etc); ultimately
it implicitly makes  assumptions about UV details,  difficult to justify without
 additional input (symmetry, etc). 
The situation can improve in models where the UV details are under control, 
such as in renormalizable models of supersymmetry breaking 
(O'Raifeartaigh, etc), not considered here (where in 
the sgoldstino decoupling limit $\Lambda$ is replaced by
an appropriate SUSY mass scale).

What does this mean for model building? When parametrizing  SUSY breaking
 in models like the MSSM one commonly 
uses a spurion field that is a limit of the goldstino superfield with the
dynamics integrated out.  The above observation regarding UV assumptions
suggests  that it may be preferable, when studying the details of a low-scale SUSY breaking case, 
to couple (offshell!) the goldstino superfield to the  MSSM, as a {\it linear}  
superfield\footnote{identified as in the text, in the case of more
sources of SUSY breaking.} rather than as a non-linear representation  that  
is a solution of the constraint  $X^2=0$. One can  then eventually decouple the  
sgoldstino explicitly, via the eqs of motion.

\newpage
\bigskip\noindent
{\bf Acknowledgements: }
The work of D. M. Ghilencea was supported by a grant of the Romanian 
National Authority for Scientific Research, CNCS - UEFISCDI, project number 
PN-II-ID-PCE-2011-3-0607.
This work  was supported in part by the European Commission  under the ERC Advanced 
Grant 226371 and the contract PITN-GA-2009-237920.

\end{document}